\documentstyle[12pt]{article}
\newcommand{\ugg}{{\; = \;}}

\newcommand{\arm}{{\rm a}}
\newcommand{\brm}{{\rm b}}
\newcommand{\crm}{{\rm c}}
\newcommand{\drm}{{\rm d}}

\newcommand{\grm}{{\rm g}}
\newcommand{\mrm}{{\rm m}}
\newcommand{\rrm}{{\rm r}}
\newcommand{\wrm}{{\rm w}}
\newcommand{\inrm}{{\rm in}}

\newcommand{\bb}{\begin{equation}}
\newcommand{\ee}{\end{equation}}
\newcommand{\bega}{\begin{eqnarray}}
\newcommand{\ega}{\end{eqnarray}}
\newcommand{\begae}{\begin{eqnarray*}}
\newcommand{\egae}{\end{eqnarray*}}

\newcommand{\h}{\hspace*{4ex}}

\newcommand{\om}{\omega}

\begin{document}

\baselineskip 0.65cm

\begin{center}
{\large {\bf ON THE PROPAGATION SPEED OF EVANESCENT MODES.}$^{\: (\dag)}$} 
\footnotetext{$^{\: (\dag)}$ Work partially supported by CAPES, and by INFN,
MURST.}
\end{center}

\vspace*{5mm}

\centerline{\ A. Pablo L. Barbero,$^{{\rm a},{\rm b}}$ Hugo E. 
Hern\'andez-Figueroa,$^\arm$ and Erasmo Recami.$^{{\rm a},{\rm c},{\rm d}}$ }

\vspace*{0.5 cm}

\centerline{{\em $^\arm$ DMO, FEEC, Universidade Estadual de Campinas, SP,
Brazil.}} \centerline{{\em $^\brm$ T.E.T., CTC, Universidade Federal Fluminense, 
RJ, Brazil.}} \centerline{{\em $^\crm$ C.C.S., Universidade Estadual de Campinas,
SP, Brazil.}} \centerline{\em $^\drm$ Facolt\`a di Ingegneria,
Universit\`a Statale di Bergamo, Dalmine (BG), Italy;} \centerline{{\em {\rm
and} INFN--Sezione di Milano, Milan, Italy.}}

\vspace*{1. cm}

{\bf Abstract \ --} \ The group-velocity of evanescent waves (in undersized
waveguides, for instance) was theoretically predicted, and has been
experimentally verified, to be Superluminal ($v_\grm > c$). By contrast, it
is known that the precursor speed in vacuum cannot be larger than $c$. \ In
this paper, by computer simulations based on Maxwell equations only, we show
the existence of both phenomena. In other words, we verify the actual
possibility of Superluminal group velocities, without violating the
so-called (naive) Einstein causality.\\

PACS nos.: \ 73.40Gk, \ 03.80+z, \ 03.65Bz \hfill\break

Keywords: evanescent waves; tunnelling photons; Hartman effect; group
velocity; Superluminal waves; precursors; transient waves; front velocity;
Maxwell equations; electromagnetic waves; computer simulations; Special 
Relativity; Extended Relativity

\newpage

{\bf 1. -- Introduction}\\

\h A series of recent experiments, performed at Cologne[1],
Berkeley[2], Florence[3] and Vienna[4], revealed that evanescent waves seem
to travel with a Superluminal group velocity ($v_\grm > c$). This originated
a lot of discussion, since it is known ---on the other hand--- that the speed
of the precursors cannot be larger than $c$. For instance, the existence of
Sommerfeld's and Brillouin's precursors (the so-called first and
second precursors) has been recently stressed in refs.[5], while studying
the transients in metallic waveguides.

\h In this paper we would like to address simultaneously both such
problems, relevant for the understanding of the propagation of a signal;
namely, the question of the (Superluminal) value of $v_\grm$ in the
evanescent case, and the question of the arrival time of the transients
(which implies a nonviolation of the so-called Einstein causality).

\h From a historical point of view, let us recall that for long
time the topic of the electromagnetic wave propagation velocity was regarded
as already settled down by the works of Sommerfeld[6] and Brillouin[7]. \ Some
authors, however, studying the propagation of light pulses in anomalous
dispersion (absorbing) media both theoretically[8] and experimentally[9],
found their envelope speed to be the group velocity $v_\grm$, even when $
v_\grm$ exceeds $c$, equals $\pm \infty$, or becomes negative! \ In the
meantime, evanescent waves were predicted[10] to be faster-than-light just
on the basis of Special Relativistic considerations.

\h But evanescent waves in suitable (``undersized") waveguides,
in particular, can be regarded also as tunnelling photons[11], due to the
known formal analogies[12] between the Schroedinger equation in presence of
a potential barrier and the Helmholtz equation for a wave-guided beam. And
it was known since long that tunnelling particles (wave packets) can move
with Superluminal group velocities inside opaque barriers[13]; therefore,
even from the quantum theoretical point of view, it was expected[13,11,10]
that evanescent waves could be Superluminal.

\h In Sect.2 of this paper we shall first show how the first
electric perturbation, reaching any point $P$, always travels with the speed 
$c$ of light in vacuum, {\em independently of the medium}. Some comments will 
be added about the instant of appearance, and the behaviour in time, of the
Sommerfeld's and Brillouin's precursors. The results of a computer
simulation will be presented for free propagation in a dispersive medium,
with the precursors arriving before the (properly said) signal.

\h In Sect.3, however, we shall deal by further computer
simulations (always based on Maxwell equations only) with evanescent 
guided-waves, showing their group velocity to be Superluminal.

\h Finally, in Sects.4 and 5 we shall deal with the transients
associated with Superluminal evanescent waves: a study that, to our
knowledge, was not carried on in the past.

\ 

{\bf 2. -- Precursors and Causality}\\

\h Every perturbation passes through a transient state before
reaching the stationary regime. This happens also when transmitting any kind
of wave. In the case of electromagnetic waves, such a transient state is
associated with the propagation of precursors, arriving before the principal
signal. This fact seems to be enough to satisfy the requirements of the
naive ``Einstein causality".

\h In particular, when investigating the {\em free} propagation
of an electromagnetic wave, in a dispersive medium with resonances in
correspondence with some
discrete angular-frequencies $\omega_j$, we can easily observe the arrival
of the first and second precursors, followed by the arrival of the properly
said signal. \ Let us consider for instance the motion in the $z$
direction of a harmonic beam, such that at $z=0$ one has:

\ 

\hfill{$f(0,t) {\; = \;} \displaystyle{{\frac{1 }{{2\pi}}} \int^{\gamma + i
\infty}_{\gamma - i \infty} {\frac{{\rm e}^{i \omega t} }{{s + i \omega}}} 
{\rm d} s} {\; = \;} {\rm e}^{i \omega t} \ {\rm for} \ t \geq 0
$\hfill} (1)

\ 

and $f(0,t) = 0$ for $t < 0$; \ where $s$ is the complex integration variable, 
and $\gamma > 0$ in order
that the function be transformable. \ Let us then consider a dispersive
medium whose dielectric constant $\varepsilon$ (electric permittivity) as a
function of $\om$ is

\

\hfill{$\varepsilon (\omega) {\; = \;} 1 + \displaystyle{\sum^N_{j=0} {\frac{
{a_j^2} }{{s^2+sg_j+\overline{\omega}_j^2}}}} \ . $\hfill} (2)

\ 

In the present model (initially proposed by Maxwell himself) $a_j$ is
proportional to the number of oscillators per unit volume, $g_j$ is the
dissipation constant (due to molecular collisions) and $\overline{\omega}
_j^2 \equiv \omega_j^2 - {\frac{1 }{3}}a_j^2$, quantity $\omega_j$ being the 
$j$-th resonant angular-frequency[14]. \ The wave equation

\[
{\frac{{\partial^2 f} }{{\partial z^2}}} - {\frac{\varepsilon }{c^2}}{\frac{{
\partial^2 f} }{{\partial t^2}}} {\ = \ } 0
\]

\ 

admits solutions of the form \ $\exp [s(t-\beta z / c)]$, \ with \ $\beta =
\beta(\omega) = \sqrt{\varepsilon(\omega)}$, \ so that we can write [for $
\gamma$ and $t$ positive]:

\ 

\hfill{$f(z,t) {\ = \ } \displaystyle{{\frac{1 }{{2\pi}}} {\displaystyle 
\int^{\gamma + i \infty}_{\gamma - i \infty}} \ \; {\frac{{\exp [s(t-\beta {\frac{
z }{c}})]} }{{s + i \omega}}} \; {\rm d} s} \ . $\hfill} (3)

\ 

Since $\beta(\omega) \rightarrow 1$ when $s \rightarrow \pm \infty$, one has
to distinguish the case $t-{z/c} < 0$ from the case $t-{z/c} > 0$. \ In the
former, the integration path in the complex plane can be closed along an
infinite-radius semicircunference on the right side, where no singularities
exist, and the integral yields zero. In other words, one gets $f(z,t) = 0$
for $t < {z/c}$, in agreement with Einstein causality. \ \
In the latter case, to look for the mentioned precursors, one has to
evaluate expression~(3) for $t-{z/c} > 0$; \ this can be accomplished by 
applying the stationary phase[15] method (which provides an illuminating 
understanding of the question), following e.g. Brillouin's[7] and 
Jackson's[16] books. For
example, the first precursor starts arriving at $t = z/c$ as a very high
frequency disturbance which grows in amplitude but decreases in frequency
with time. Its amplitude, after the maximum, decreases till the arrival of
the second precursor, which $\;$---when there is only a resonance ($j=0$) at $
\omega = \omega_0$, and $g = 0$---$\;$ starts at time \ $t {\; = \;} z \sqrt{(
\overline{\omega}_0^2 + a^2)/{\overline{\omega}_0^2}}/c$, \ reaches a
maximum, and then decreases, while the oscillation angular-frequency tends
to the initial excitation angular-frequency $\omega$ which enters eq.(1). The
properly said signal arrives afterwards (independently of the medium).

\h If we pass to consider, however, non-free propagation (in the
vacuum) inside a wave guide, when a cutoff angular-frequency $\omega_\crm$
enters the play, the stationary phase method application is restricted by
the fact that the propagation constant \ $\beta(\omega) {\; = \;} \omega 
\sqrt{1-(\omega_\crm/\omega)^2}/c$ \ becomes imaginary for $\omega <
\omega_\crm$. \ Nevertheless, if the beam contains also above-cutoff
spectral components, then the first precursor evaluation ---which depends
only on the highest frequencies--- are still possible, as shown, e.g., by
Stenius and York[5]. \ We shall discuss such problems in the next
Section.

\h Here, let us just simulate the free propagation of an
electromagnetic field in a medium described by eq.(2) with $j=0$, i.e., 
described by

\ 

\hfill{$\varepsilon (\omega) {\; = \;} 1 + \displaystyle{{\frac{{a_0^2} }{{
\overline{\omega}_0^2-\omega^2-i\omega g_0}}}} \ , $\hfill} (2')

\ 

with $a_0 = 2.2 \times 10^{10}, \ \omega_0 = 4.4 \pi \times 10^{10}$, and $
g_0 = -10^9$. Let us assume the electric field at $z=0$ to be \ $f(0,t) {\;
= \;} A \; t \; \exp[-at^2] \; \sin(\omega t)$ \ with $A=10^9$ and $a=5
\times 10^{17}$. Fig.1 shows such a function of time (in ns) for $\omega =
7$ GHz. \ The calculations then yield, for $z = 63$ m, the electric
field in Fig.2. \ For evidencing the Sommerfeld and Brillouin precursors, it is
necessary to magnify the vertical scale by a factor $10^4$: see Fig.3, where
the horizontal axis is still the time axis (in ns). \ Fig.3 shows that the 
electric
perturbation starts at $t=210$ ns, corresponding to the time needed to
travel 63 m with speed $c$, when the first precursor starts arriving at $z$
as a very high frequency field (in fact, the stationary phase method expects
that immediately after $t=z/c$ only the highest frequency components
contribute to the integral (3)). \ The second precursor starts reaching $z$
at $t \simeq 212.6$ ns, in perfect agreement ---again--- with the stationary 
phase solution. \ Afterwards, the field angular-frequency tends to 
$\omega = 7$ GHz (stationary regime) and the properly said signal
starts arriving.

\ 

{\bf 3. -- Propagation below the cutoff frequency}\\

\h Let us come to the point we are more interested in, i.e., to
the propagation in waveguides of pulses obtained by amplitude modulation
of a carrier-wave endowed with an under-cutoff frequency; and let us recall 
that the experiments ---for instance--- in
refs.[1-4] did actually detect in such a case a Superluminal group-velocity, 
$v_\grm > c \;$ (in agreement with the classical[10] and the quantum[13]
predictions).

\h For example, the work in refs.[1,17] put in particular
evidence the fact that the segment of ``undersized" (= operating with
under-cutoff frequencies) waveguide provokes an attenuation of each spectral
component, without any phase variation. More precisely, the unique phase
variation detectable is due to the discontinuities in the waveguide
cross-section (cf. also refs.[13]). Mathematically[18], the spectrum leaving
an undersized waveguide segment (or {\em photonic barrier\/}) is simply the
entering spectrum multiplied by the transfer function \ $H(\omega) {\; =
\;} \exp[i\beta L]$, \ with \ $\beta(\omega) {\; = \;} \omega \sqrt{
1-(\omega_\crm/\omega)^2}/c$. \ For $\omega > \omega_\crm$, the propagation
constant $\beta(\omega)$ is real, and $H(\omega)$ represents a phase
variation to be added to the outgoing spectrum. \ However, for $\omega <
\omega_\crm$, when $\beta(\omega)$ is imaginary, the transfer function
just represents an additional attenuation of the incoming spectrum.

\h In a sense, the two edges of a ``barrier" (undersized waveguide segment: see
Fig.4) can be regarded as semi-mirrors of a Fabry--Perot configuration. The
consequent negative interference processes can lead themselves to
Superluminal transit times. This points have been exploited, e.g., by Japha 
and Kurizki[19] (who claimed the barrier transit mean-time to be Superluminal
provided that the coherence time $\tau_\crm$ of the entering field
$\psi_\inrm(t)$ is much larger than $L/c$).

\ 

{\bf 4. -- Our numerical experiments}\\

\h As already mentioned, to investigate the interplay between
Einstein causality and the fact that $v_\grm \gg c$ when a signal is
transported in a metallic waveguide by a carrier-wave with $\omega_\wrm <
\omega_\crm$, one has to examine simultaneously the effects mentioned in
Sects.2 and 3.

\h Let us consider a signal obtained by a pulse-shaped amplitude
modulation of a carrier-wave with frequency $f_\wrm$ (in Fig.5 the envelope
of the wave is shown). \ Let us assume that the carrier-wave is switched on
at time $t=0$, so that at the (undersized) waveguide entrance ($z=0$) the
field will be \ $f(0,t) = 0$ \ for $t < 0$. The amplitude of the
carrier-wave will reach a stationary state soon after the rise-time instant, 
$t_\rrm$ (here defined as the time requested for the carrier amplitude to
increase from 10\% to 90\% of its stationary value). \ A (smoothly prepared) 
gaussian pulse, with
width $\Delta t$, be centered at $t = t_\mrm, \ \ (t_\mrm > t_\rrm)$. \ At
time $t = t_\drm, \ \ (t_\drm > t_\mrm + \Delta t)$, the carrier wave is
switched off (and its amplitude will decrease in a time of the order of $
t_\rrm$). \ Wishing to reveal the precursors too, it is important to use
values of $t_\rrm$ smaller than 100 ps (so to excite the higher frequency
components with enough power).  \ It is important, as well, to use a
spectrally narrow pulse ($\Delta \omega \ll \omega_\wrm$), so that one can
go on calculating the group-velocity via the standard relation $v_\grm = {
\partial \omega}/{\partial \beta}$.

\h A spectrally narrow pulse, moreover, allows us to examine the 
{\em double barrier} experiment[20], i.e. the most interesting
configuration, without making recourse to external filters. The setup is
shown in Fig.6; 
the {\em two} photonic {\em barriers} are segments of undersized waveguide
25 and 50 mm long, respectively, with cross-section $23.45 \times 34.85$ mm$
^2$ and cutoff frequency 4.304 GHz. Between them, there is another segment,
101 mm long, of ``normal-sized" waveguide, with cross-section $23.45 \times
48.85$ mm$^2$ and cutoff frequency 3.07 GHz. \ The transfer function,
illustrated in Fig.7, was calculated by using a Fortran program[21] based on 
the method of moments (MoM), while the mode
decomposition was performed in terms even modes TE$_{m0}$, with $m$ an odd
number. As usual, the outgoing spectrum was evaluated by multiplying the
incoming spectrum (Fig.5) by the transfer function, that is to say by
use of the inverse Fourier transform (within the software package 
{\em Mathematica} 2.2.3). \ It was chosen a carrier-wave with
frequency $f_\wrm = 3.574$ GHz, corresponding to a minimum of $\partial
\phi / \partial \nu$, where $\phi$ is the transfer-function phase. Let
us recall that the magnitude of the transfer function for this frequency
is the attenuation suffered by the electromagnetic wave along the two
photonic barriers. \ The outgoing electric signal is shown in Fig.8; in its
inset (a) one can see the exact arrival time $t \simeq 0.488$ ns, at the
exit interface, of the {\em first} electric disturbance (such an instant differing
a little from the one, $t = L/c \simeq 0.587$ ns, predicted in Sect.2, since
in our simulation we used of course a finite ``sample rate", 0.4884 ns; by
reducing ths rate, a better result is obtained). \ In inset (b) 
we see the entering gaussian
pulse, initially modulated and centered at $t = 800$ ns.

\h In Fig.9(a) the pulse peak is represented in more detail.  From its arrival
time, $t \simeq 800.24$ ns, we can derive the ({\em Superluminal\/})
group-velocity $v_\grm {\; = \;} (176/0.24)$ mm/ns $\simeq \; 7.33 \times
10^8$ m/s $\simeq \; 2.44 \; c$. \ If we want to evaluate the group-velocity
by the relation $v_\grm {\; = \;} {\partial \omega}/{\partial \beta}$, we
get (all the derivatives being evaluated at the frequency $f_\wrm$ of the
carrier-wave):

\ 

\hfill{$v_\grm {\ = \ } \displaystyle{\left.{\frac{{\partial \omega} }{{
\partial \beta}}}\right|_{f_\wrm} {\ = \ } 2\pi \, \left.{\frac{{\partial \nu
} }{{\partial \beta}}}\right|_{f_\wrm} {\ = \ } \frac{2\pi}{\left. {\frac{{
1000} }{{176}}} \, {\frac{{\partial \phi} }{{\partial \nu}}}\right|_{f_\wrm}}
} \ \simeq \ 2.48 \; c \ , $\hfill} (4)

\ 

in very good agreement with the previous value (their difference being smaller 
than 2\%). \ In the previous simulation we used a pulse half-width $\Delta 
\nu { \; = \;} 12$ MHz, so that, as required, ${\Delta \nu}/{f_\wrm} \simeq 
0.0034 \ll 1$.

\h Notice that the $0.24$ ns spent by the pulse inside the setup
of Fig.6 is due to the wave phase variation caused by the
geometric discontinuities existing between the different waveguide segments
which compose the analyzed setup (mainly the leading edges of the
barriers): we shall come back to this point. \ One can therefore expect[22]
such a transit time to be {\em independent} not only of the length of the
barriers ({\em Hartman effect\/}: see refs.[13]), but even {\em of the
length of the ``normal" waveguide inserted between the two barriers}. \ This
has been experimentally verified[20], and constitutes the most interesting
fact revealed by refs.[1,17,20].

We repeated our computer simulation for the same setup depicted in Fig.6,
when inserting between the undersized waveguides (barriers) a segment of
``normal" waveguide $501$ mm (instead of $101$ mm) long; with a new,
suitable choice of the carrier frequency ($f_\wrm = 3.5795$ GHz). \ The new
pulse can be seen in Fig.9(b). \ The delay (transit time) resulted to be $
0.336$ ns, corresponding to a {\em higher (Superluminal)} group-velocity, \ $
v_\grm {\; = \;} (576/0.336)$ mm/ns $\simeq \; 17.14 \times 10^8$ m/s $
\simeq \; 5.71 \; c$. \ Again, by using the standard definition, we obtain
the very close value

\hfill{$v_\grm {\ = \ } \displaystyle{\left.{\frac{{\partial \omega} }{{
\partial \beta}}}\right|_{f_\wrm}} \ \simeq \ 5.91 \; c \ , $\hfill} (4')

\ 

their difference being less than 3.4\%. \ Let us notice that the considered
setup (Fig.6) works as a Fabry--Perot filter, so that, when the length $L_2$
of the intermediate (``normal-sized") waveguide increases, the usable band
width decreases. \ Of course, if we had chosen a carrier frequency outside
the suited intervals, e.g. $f_\wrm {\; = \;} 5.58945$ GHz ({\em non}-evanescent
case), we would have got a subluminal group-velocity.  In fact, our
calculations yield in this case that the outgoing pulse (see Fig.9c) is
centered at $t {\; = \;} 0.977$ ns, corresponding to the group-velocity $
v_\grm {\; = \;} (176/0.977)$ mm/ns $\simeq \; 0.6 \; c$.

\ 

{\bf 5. -- The case of an infinite undersized waveguide.}\\

\h Let us stress once more that all the delays (non-zero transit
times) found above, in our simulations of experiments, depend only on the
phase variation suffered by the wave because of the geometric
discontinuities in the waveguide. Actually, as already mentioned, the 
propagation constant $\beta
(\omega )$ is imaginary for the under-cutoff frequencies, so that the
transfer function $H(\omega )$ works only as an attenuation factor for
such (evanescent) frequencies. However, the higher (non-evanescent)
frequencies will be phase shifted, in such a way that $\beta (\omega )$ will
tend to its free-space value $\omega /c$ for $\omega \rightarrow \infty $.
In other words, the higher spectral components travel with speed $c$; they
are the responsible both for the finite speed of the evanescent beams, and
for the appearance of the precursors. \ [In the (theoretical) case that a
pulse were constituted by under-cutoff frequencies only, the situation could
therefore be rather different]. 

\h Anyway, let us eliminate the effect of the geometric
discontinuities just by considering an electromagnetic {\em signal} which is
already propagating inside an under-sized waveguide, and travelling between
two parallel cross-sections separated by the distance $L$. \ The waveguide
size be $5 \times 10$ mm$^2$, and $L = 32.96$ mm (cf.Fig.10). \ The entering
signal envelope is shown in Fig.11 as a function of time; the (smoothly
prepared) gaussian
pulses are centered at $t_\mrm {\; = \;} 100, \ 170, \ 240$ and $300$ ns,
respectively. \ In inset (a) the initial part (in time) of the mentioned
envelope is shown, while in inset (b) one can see the peak of the gaussian 
pulse
centered at 100 ns. \ After having travelled the considered distance $L$
through the undersized waveguide (characterized by the transfer function
depicted in Fig.12), the evanescent signal arrives with the envelope shown
in Fig.13. The shape is essentially the same (cf. also inset (b) of Fig.13),
even if the amplitude is of course reduced. \ In inset (a) of Fig.13 one can
see the {\em initial} part (in time) of the transmitted signal, arriving after
109.87 ps, which is exactly the time needed to travel 32.96 mm with the
speed $c$ of light in vacuum. \ However, by comparing insets (b) of Figs.11
and 13, one deduces that the pulses travelled with {\em infinite}
group-velocity, since the transmission of the pulse-peaks required zero time
(instantaneous transmission).

\h It is interesting also to analyze the spectra of the entering (Fig.14) and 
arriving (Fig.15) signal. Fig.14 shows the Fourier transform of the signal 
presented in Fig.11, when it modulates in amplitude a carrier-wave with 
frequency 14.5 GHz.
\ In the insets of Figs.14 and 15, we show the signal spectrum after
magnifying the vertical scale by a factor $3\times 10^4$; we can notice that
the arriving signal possesses a spectral component (approximately centered
at 15 GHz) that was not present in the entering spectrum: such a new
component corresponds to the waveguide cutoff value, 15 GHz in this case. 

\ After the transients, the
real signal arrives, with a Superluminal (even infinite) group-velocity.

\ 

{\bf 6. -- Conclusions.}\\

\h At this point, one can accept that a signal is really carried (not by 
the precursors, but) by well-defined amplitude bumps, as in the case of
information transmission by the Morse alphabet, or the transmission of a
number e.g. by a series of equal (and equally spaced) pulses. \ In such a case,
we saw above that the signal can travel even at infinite speed,
in the considered situations.  It is important also to notice, when comparing
Fig.13 with Fig.11, that the width of the arriving pulses does not change with 
respect to the initial ones. \ The signal, however, cannot overcome the
transients, ``slowly'' travelling with speed $c$.

\h Even if the AM signal were totally constituted by under-cutoff frequencies, 
when the experiment is started (e.g., by switching on the carrier wave) one
does necessarily meet a transient situation, which generates precursors.

\h One might think, therefore, of arranging a setup (permanently switched on) 
for which the precursors are sent out long in advance, and waiting afterwards 
for the moment at which the need arises of transmitting a signal
with Superluminal speed (without violating the naive ``Einstein causality",
as far as it requires only that the precursors do not travel at speed higher 
than $c$). \ Some authors, as the ones in refs.[1,17,20], do actually claim 
that they can build up (smooth) signals by means of under-cutoff frequencies 
only, {\em without generating further precursors\/}: in such a case one 
would be in presence, then, of Superluminal information transmission.

\h However, on the basis of our calculations (which imply the existence also 
of above-cutoff frequencies in any signal: cf. the inset of Fig.14) this does 
not seem to be true in practice. \ If, in reality, to start sending out a 
signal means to create some discontinuities (i.e., to generate new precursors),
and if the signal cannot bypass the precursors (even when the carrier was
switched on long in advance), then information could not be transmitted 
faster than light by the experimental devices considered above, in spite of 
the fact that evanescent signals travel with Superluminal group-velocity.  

\h Such critical issues deserve further investigation, and we shall come back 
to them elsewhere (for instance, a problem is whether one must already know 
the whole information content of the signal when {\em starting} to send it;
in such a case, it would become acceptable the mathematical trick of 
representing any signal by an analytical function[23]). \ But we have seen 
that, in any case, the evanescent modes travel for some distance with 
faster-than-light speed; and at least in
three further sectors of experimental physics Superluminal motions {\em might}
have been already observed[24]. \ Therefore, it is worthwhile to recall here, 
in this regard, that Special Relativity 
{\em itself} can, and was, extended[25] to include also Superluminal motions 
on the basis of its {\em ordinary} postulates; solving seemingly also the 
known causal paradoxes[26] associated in the past with tachyonic motions.\\  

\ 

\ 

\ 

ACKNOWLEDGMENTS\\ 

We thank the Antennas Group of the Telebr\'as Research Center in Campinas, SP, 
Brazil, for allowing us to use their MoM code. \  For stimulating discussions, 
we are grateful to R.Bonifacio, R.Chiao, A.Del Popolo, R.Garavaglia, 
D.Jaroszynski, 
L.C.Kretly, G.Kurizki, G.Nimtz, A.Steinberg, J.W.Swart and M.Zamboni-Rached.
At last, for providing us with kind constant computer assistance, we
thank also Professor Y.Akebo \ and \ Rodrigo L.Anami, \ Adriano Domingos N., \
V\'{\i}taly F.Rodr\'{\i}gues E., \ Janette Toma.\\

FIGURE CAPTIONS:\\

Fig.1 -- The electric field at $z = 0$ as a function of time (in ns), for
$\om = 7$ GHz (see the text).\\

Fig.2 -- The same electric field considered in Fig.1, after having travelled
63 m in a medium characterized by eq.(2').\\

Fig.3 -- Same as Fig.2, with the vertical scale magnified by a factor $10^4$. 
The Sommerfeld and Brillouin precursors start arriving at times $t_0$ and 
$t_1$, respectively.\\

Fig.4 -- A waveguide with a segment of ``photonic barrier", i.e., of 
undersized waveguide (evanescence region).\\

Fig.5 -- Envelope of a gaussian signal (centered at $t_\mrm \ugg 800.00$ ns, 
with width $\Delta t \ugg 37.32$ ns) obtained by amplitude modulation of a 
carrier-wave. \ We assume the carrier-wave to be switched on at time $t = 0$;
inset (a) shows the rise time, $t_\rrm \ugg 37.70$ ns, of the carrier 
amplitude (for increasing from 10\% to 90\% of its stationary value).\\

Fig.6 -- The experimental setup considered for our simulations.\\

Fig.7 -- The transfer function corresponding to the setup in Fig.6.  Its
magnitude and phase are represented by the pointed and solid
lines, respectively. 
Notice that the intervals in which the phase derivative is lower coincide
with the dips of the magnitude.\\

Fig.8 -- Aspect of the signal in Fig.5, after having propagated through the
setup in Fig.6.  Inset (a) shows the arrival time of its initial part.\\

Fig.9 -- Detailed representation of the signal peak, after propagation 
through the setup in Fig.6 with different lengths $L_2$ of the intermediate
(``normal-sized") waveguide and with different carrier frequencies $f_\wrm$: \ 
(a) $L_2 = 101$ mm, and 5.574 GHz; \ (b) $L_2 = 501$ mm, and $3.5795$ GHz; \ (c) 
again $L_2 = 101$ mm, but $f_\wrm = 5.58945$ GHz.\\

Fig.10 -- The (indefinite) {\em undersized} waveguide considered in our simulations,
when eliminating any geometric discontinuity in its cross-section.  We chose
$L = 32.96$ mm.\\

Fig.11 -- Envelope of the initial signal, considered in our simulation for
signal propagation through the new setup in Fig.10. \ Inset (a) shows in
detail the initial part of this signal as a function of time, while inset (b)
shows the gaussian pulse peak centered at $t = 100$ ns.\\

Fig.12 -- The transfer function corresponding to the new setup in Fig.10. \
Its magnitude and phase are represented by lines (a) and (b), respectively.\\

Fig.13 -- Envelope of the signal in Fig.11 after having propagated through
the undersized waveguide in Fig.10. \ Inset (a) shows in detail the initial
part (in time) of such arriving signal, while inset (b) shows the peak of
the gaussian pulse that had been initially modulated by centering it at $t =
100$ ns (one can see that its propagation took {\em zero} time).\\

Fig.14 -- Spectrum of the entering signal.  In the inset, the vertical scale
was magnified $3 \times 10^4$ times.\\

Fig.15 -- Spectrum of the arriving signal. From the inset, where the vertical 
scale was again magnified by the factor $3 \times 10^4$, one can notice the appearance 
of a new spectral component at 15 GHz.\\

\

REFERENCES\\

[1] A.Enders and G.Nimtz: J. de Physique-I 2 (1992) 1693; 3 (1993) 1089;
Phys. Rev. E48 (1993) 632; \ G.Nimtz, A.Enders and H.Spieker: J de
Physique-I 4 (1994) 1; \ W.Heitmann and G.Nimtz: Phys. Lett. A196 (1994)
154; \ G.Nimtz: Physik Bl. 49 (1993) 1119; ``New knowledge of tunnelling
from photonic experiments", in {\em Tunneling and its Implications} (World
Scient.; Singapore, in press); \ G.Nimtz and W.Heitmann: ``Photonic bands
and tunneling", in {\em Advances in Quantum Phenomena}, ed. by
E.G.Beltrametti and J.-M.L\'evy-Leblond (Plenum Press; New York, 1995),
p.185; \ Prog. Quant. Electr. 21 (1997) 81; \ H.Aichmann and G.Nimtz:
``Tunnelling of a FM-Signal: Mozart 40", submitted for pub. \ See also
refs.[20]. \ [Nimtz et al. made also same computer simulations (on the basis
of Maxwell eqs.), well reproducing the related experimental results: see 
ref.[17].]\hfill\break

[2] A.M.Steinberg, P.G.Kwiat and R.Y.Chiao: Phys. Rev. Lett. 71 (1993) 708;
\ R.Y.Chiao, P.G.Kwiat and A.M.Steinberg: Scientific American 269 (1993),
issue no.2, p.38. \ Cf. also A.M.Steinberg and R.Y.Chiao: Phys. Rev. A51
(1995) 3525; P.G.Kwiat et al.: Phys. Rev. A48 (1993) R867; \ E.L.Bolda et
al.: Phys. Rev. A48 (1993) 3890.\hfill\break

[3] A.Ranfagni, P.Fabeni, G.P.Pazzi and D.Mugnai: Phys. Rev. E48 (1993)
1453. \ Cf. also Appl. Phys. Lett. 58 (1991) 774.\hfill\break

[4] Ch.Spielmann, R.Szipocs, A.Stingl and F.Krausz: Phys. Rev. Lett. 73
(1994) 2308.\hfill\break

[5] See, e.g., P.Stenius and B.York: IEEE Ant. and Prop. Mag.(?), 37 (1995)
39. \ Cf. also S.L.Dvorak: IEEE Trans. Microwave Th. and Techn. 42 (1994)
2164.\hfill\break

[6] A.Sommerfeld: Z. Physik 8 (1907) 841.

[7] L.Brillouin: Ann. Physik 44 (1914) 203; \ {Wave Propagation and Group
Velocity} (Academic Press; New York, 1969).\hfill\break

[8] C.G.B.Garrett and D.E.McCumber: Phys. Rev. A1 (1970) 305.\hfill\break

[9] S.Chu and S.Wong: Phys. Rev. Lett. 48 (1982) 738. \ See also M.W.Mitchell
and R.Y.Chiao: Phys. Lett. A230 (1997) 133.\hfill\break

[10] Cf., e.g., pages 158, and 116-117, in E.Recami: Rivista Nuovo Cim. 9
(1986), issue no.6, pp.1-178, and references therein.\hfill\break

[11] S.Bosanac: Phys. Rev. A28 (1983) 577.\hfill\break

[12] See, e.g., Th.Martin and R.Landauer: {\em Phys. Rev. A\/}{\bf 45}
(1992) 2611; \ R.Y.Chiao, P.G.Kwiat and A.M.Steinberg: {\em Physica\/}B{\bf 
175} (1991) 257; \ A.Ranfagni, D.Mugnai, P.Fabeni and G.P.Pazzi: {\em Appl.
Phys. Lett.} {\bf 58} (1991) 774. \ See also A.M.Steinberg: {\em Phys. Rev.}
A52 (1995) 32.\hfill\break

[13] See V.S.Olkhovsky and E.Recami: Phys. Reports 214 (1992) 339, and refs.
therein; in particular T.E.Hartman: J. Appl. Phys. 33 (1962) 3427;
J.R.Fletcher: J. Phys. C18 (1985) L55; F.E.Low and P.F.Mende: Ann. of Phys.
210 (1991) 380; V.S.Olkhovsky, E.Recami, F.Raciti and A.K.Zaichenko: J. de
Physique-I 5 (1995) 1351; D.Mugnai et al.: Phys. Lett. A209 (1995) 
227-234.\hfill\break

[14] See, e.g., A.Stratton: ``Electromagnetic Theory" (McGraw-Hill; New
York, 1941), p.322.\hfill\break

[15] E.P.Wigner: Phys. Rev. 98 (1955) 145.\hfill\break

[16] D.Jackson: ``Classical Electrodynamics" (New York, 1975).\hfill\break

[17] H.M.Brodowsky, W.Heitmann and G.Nimtz: Phys. Lett. A222 (1996) 125.\hfill\break

[18] M.Schwartz: ``Information Transmission, Modulation and Noise"
(McGraw-Hill; New York, 1970).\hfill\break

[19] Y.Japha and G.Kurizki: Phys. Rev. A53 (1996) 586. \ Cf. also G.Kurizki,
A.Kozhekin and A.G.Kofman: Europhys. Lett. 42 (1998) 499: \ G.Kurizki,
A.E.Kozhekin, A.G.Kofman and M.Blaauboer: presented at the VII Seminar on
Quantum Optics, Raubichi, BELARUS (May, 1998).\hfill\break

[20] G.Nimtz, A.Enders and H.Spieker: J. de Physique-I 4 (1994) 565; \
``Photonic tunnelling experiments: Superluminal tunnelling", in {\it Wave
and Particle in Light and Matter -- Proceedings of the Trani Workshop,
Italy, Sept.1992}, ed. by A.van der Merwe and A.Garuccio (Plenum; New York,
1993); \ A.Enders and G.Nimtz: Phys. Rev. B47 (1993) 9605.\hfill\break

[21] Numerical code developed by the Antennas Group of the Telebr\'as
Research Center, Campinas, SP, Brazil.\hfill\break

[22] J.Jakiel, V.S.Olkhovsky and E.Recami: to appear in Phys. Lett. A; \
V.S.Olkhovsky, E.Recami and J.Jakiel: ``Unified approach to the
tunnelling time for particles and photons" (submitted for pub.); \
V.S.Olkhovsky and A.Agresti: in {\em Tunneling and its Implication}, ed.
by D.Mugnai, A.Ranfagni and L.S.Schulman (World Scient.; Singapore, 1997), 
pp.327-355; \ V.S.Olkhovsky: Fizika Zhivogo (Physics of the Alive) 5 (1997)
19-37; \ V.S.Olkhovsky, E.Recami and A.Agresti: ``New developments in the study
of Time as a quantum observable", to be submitted for pub.\hfill\break

[23] Cf., e.g., F.E.Low: (private communication).\hfill\break

[24] See, e.g., E.Recami: ``On localized 'X-shaped' Superluminal solutions 
to Maxwell equations", {\em Physica A}{\bf 252} (1998) 586-610, and references
therein. \ See also E.Recami: ``Some information about the four experimental
sectors of physics in which Superluminal motions seem to appear", Report
INFN/FM-97/01 (INFN; Frascati, 1997).\hfill\break

[25] See, e.g., E.Recami: ``Classical tachyons, and possible applications",
Rivista N. Cim. 9 (1986), issue no.6 (pp.1-178), and refs. therein.\hfill\break

[26] See, e.g., E.Recami: ``Tachyon Mechanics and Causality; A Systematic
Thorough Analysis of the Tachyon Causal Paradoxes", Foundations of Physics
17 (1987) 239-296.\hfill\break

\end{document}